\documentclass[10pt,a4paper]{article}
\usepackage{t1enc}
\usepackage[final]{graphicx}
\usepackage{graphicx}
\usepackage{epsfig}

\usepackage{mathrsfs}
\usepackage{amsmath}
\usepackage{amsfonts}
\usepackage{amssymb}
\numberwithin{equation}{section}
\usepackage{latexsym}
\usepackage{amsfonts}
\usepackage{amsmath}
\usepackage{amsmath,amssymb,calc,amsfonts}
\usepackage{latexsym}
\usepackage{graphicx,calc,epsfig}
\usepackage{t1enc}
\usepackage{times}
\usepackage{xmpmulti}
\usepackage{color}
\usepackage{bm}

\font\tenscr=rsfs10 scaled1100
\font\sevenscr=rsfs7 
\font\fivescr=rsfs5 
\skewchar\tenscr='177
\skewchar\sevenscr='177
\skewchar\fivescr='177
\newfam\scrfam
\textfont\scrfam=\tenscr
\scriptfont\scrfam=\sevenscr
\scriptscriptfont\scrfam=\fivescr

\def\scri{{\fam\scrfam I}}

\begin{document}
\title{Spin and Center of Mass in Axially Symmetric Einstein-Maxwell Spacetimes}
\author{
\small C. N. Kozameh \\
\em \small FaMAF, Universidad Nacional de C\'ordoba\\
\normalsize \em \small 5000, C\'ordoba, Argentina \\
\normalsize \em \small kozameh@famaf.unc.edu.ar\\
\\
\\
\small G. D. Quiroga\\
\normalsize \em \small FaMAF, Universidad Nacional de C\'ordoba\\
\normalsize \em \small 5000, C\'ordoba, Argentina \\
\normalsize \em \small gquiroga@famaf.unc.edu.ar
}
\maketitle
\begin{abstract}
We give a definition and derive the equations of motion for the center of mass and angular momentum of an axially symmetric, isolated system that emits gravitational and electromagnetic radiation. A central feature of this formulation is the use of Newman-Unti cuts at null infinity that are generated by worldlines of the spacetime.  We analyze some consequences of the results and comment on the generalization of this work to general asymptotically flat spacetimes.
\end{abstract}

\section{Introduction}
The notion of center of mass for an isolated system is very important in newtonian theory. It is used to define the linear and intrinsic angular momentum of the system, both conserved observables in the theory. However, its generalization to General Relativity (GR) has proved to be a non trivial task.

 A major obstacle to a relativistic definition is the fact that energy or momentum cannot be local quantities. One cannot use the Energy Momentum tensor of a system as in other theories since gravitational waves carry away energy and momentum and nevertheless are solutions of Ricci flat equations. Thus, one must search for global definitions of these quantities. Also one must bear in mind that, unless the spacetime is stationary, energy or momentum of an isolated system are not conserved in GR due to the emission of gravitational radiation.

Fortunately there are ways to overcome, at least in principle, these difficulties. One has available in the literature the notion of an asymptotically flat spacetime, the appropriate framework to analyze isolated systems in GR. For those spacetimes one can define the notion of Bondi mass and linear momentum and write down equations of motion linking their time evolution with the emitted gravitational radiation. All that is needed then is to relate these variables to a suitable definition of center of mass, characterized by a worldline $R^a$, such that $P^a= M \dot{R}^a +$ radiative corrections. The problem is how to select this worldline. There is available in the literature two nice approaches based on information that can be retrieved at null infinity $\scri$.

One approach that has been carried out by Newman and collaborators is based on the introduction of asymptotically shear free null congruences, i.e., congruences such that at $\scri$ have vanishing shear. At null infinity this congruence appears to come from a point in Minkowski space. This asymptotically shear free condition yields a family of two-surfaces, "good cuts", that are constructed from special solutions of the Good Cut equation. This family is characterized by complex worldlines in a fiducial holographic space. Furthermore, the vanishing of the complex mass dipole term at $\scri$ singles out a particular worldline of this family. Its real part gives the center of mass while the imaginary part is by definition the intrinsic angular momentum per unit mass. Assuming a quadrupole radiation and using any available definition of total angular momentum (where all of them coincide) one obtains equations of motion coupling center of mass, intrinsic angular momentum and radiation. (A complete description is available at a Living Reviews \cite{akn}).

The other approach has been done by Moreschi \cite{moreschi2}. In this case one first defines the notion of "supermomentum" at $\scri$ and then asks for a special family of cuts, called nice cuts, where the supermomentum only has an $l=0$ and $l=1$  spherical harmonics decomposition. Again one obtains another holographic solution space and one special worldline in this space is selected, via a similar condition as above, as the center of mass. Furthermore. Moreschi defines the notion of total angular momentum \cite{moreschi1} and its restriction to the center of mass worldline yields the intrinsic angular momentum of the system.

Both formulations coincide at a linear level if the gravitational radiation is pure quadrupole. In spite of the clever ideas used in both approaches to define a global notion of center of mass there are some drawbacks that one must mention since in the end the definition should be used in gravitational wave astronomy to characterize the dynamics of compact highly energetic objects like AGN or binary systems such as BH-BH or BH-NS.

It is not true that light coming from distant isolated systems is asymptotically shear free. In fact the shear of light coming from these sources is used to define weak lensing effects in GR. Most important, neither the nice cuts nor the good cuts are  related to the future light cones from points inside the spacetime. One can write down the equation that must be satisfied by any light cone cut at $\scri$ coming from a worldline in the spacetime \cite{BKR}. None of these approaches satisfy this equation up to second and higher orders.

The idea of this note is to use the light cone cut equation recently obtained \cite{BKR} together with ideas borrowed from the two approaches to define center of mass and intrinsic angular momentum. So far, there is not available in the literature a satisfactory definition of angular momentum for non-stationary spacetimes without symmetries. To overcome this difficulty we will consider here axially symmetric spacetimes since in this case there is a suitable definition of angular momentum.

This work is divided in five sections and two appendices. Section 2 is devoted to the mathematical tools and definitions needed for this work. Readers familiar with the Newman-Penrose formulation, asymptotic flatness and  the Komar integral may skip this section. Sections 3 and 4 constitute the core of the paper. Finally, in the Conclusions we summarize our results and outline a generalization of this formalism to arbitrary spacetimes.

\section{Foundations}
\quad  There are many result that are needed for this work. In this section, we introduce several of the key ideas and the basic
tools that are indispensable in our later discussions. A thorough derivations of these results are given in the references.
\subsection{Asymptotically flat spacetime and $\scri ^{+}$}
\quad  The notion of  asymptotically flatness is the adequate tool to analyze the gravitational and electromagnetic radiation
coming from an arbitrary compact source. A spacetime can be thought of as asymptotically flat if the curvature tensor vanishes
as infinity is approached along the future-directed null geodesics of the spacetime. All the null geodesics and up at what
is referred to as future null infinity, $\scri ^{+}$, the future boundary of the spacetime \cite{akn}, \cite{ntod}.
We introduce a natural set of coordinates in the neighborhood of $\scri ^{+}$ called Bondi coordinates
($u_{B}$, $r$, $\zeta$, $\overline{\zeta }$). In this system, the Bondi time $u_{B}$ labels a special family of null surfaces whose intersection with $\scri$ are two spheres, $r$ is the affine
parameter along each null geodesic of the constant $u_{B}$ surface and $\zeta =e^{i\varphi }\cot \frac{\theta }{2}$, is the complex
stereographic angle that labels the null geodesics of the null surface.

Associated with the Bondi coordinates is a null tetrad system based on these outgoing null hypersurfaces labeled by
($l_{a}$, $n_{a}$, $m_{a}$, $\overline{m}_{a}$). The first tetrad vector $l_{a}$ is defined as \cite{ntod}
\begin{equation}
l_{a}=\nabla _{a}u_{B}.
\end{equation}
Thus, $l^{a}=g^{ab}\nabla _{b}u_{B}$ is a null vector tangent to the geodesics of the surface.
For the second tetrad vector we pick a null vector $n^{a}$ normalized to $l^{a}$
\begin{equation} \label{ln}
n_{a}l^{a}=1
\end{equation}
The tetrad is finally completed with the choice of a complex null vector $m^{a}$ orthogonal to $l^{a}$ and $n^{a}$
\begin{equation}\label{mmb}
m_{a} \overline{m}^{a}=-1
\end{equation}
The spacetime metric is then \cite{ntod}
\begin{equation} \label{metrica}
g_{ab}=l_{a}n_{b}+n_{a}l_{b}-m_{a}\overline{m}_{b}-\overline{m}_{a}m_{b}
\end{equation}
There is a great number of tetrad freedom, but the most important for us is a different choice of the original $u_{B}=const.$ cuts of
$\scri ^{+}$ so that
\begin{equation}\label{ubz}
u_{B}=Z(u, \zeta , \overline{\zeta})
\end{equation}
where $Z(u, \zeta , \overline{\zeta})$ is a real function. Let us denote by $T$ the inverse function $Z$, so
\begin{equation}
u=T(u_B,\zeta,\overline{\zeta})
\end{equation}
is easy to show that $\dot{T}= \frac{1}{Z^{\prime}}$, then the rest of the coordinate system and the tetrad system are then constructed as before.

A second freedom introduces the concept of spin weight \cite{ntod}. A quantity $\eta$ that transforms as $\eta \rightarrow e^{is\lambda }\eta$ under a rotation $m^{a} \rightarrow e^{i\lambda }m^{a}$ is said to have a spin weight $s$. For any function $f(u, \zeta, \overline{\zeta})$, we can define two differential operators $\eth$ and $\overline{\eth}$ by
\begin{eqnarray}
\eth f&=&P^{1-s}\frac{\partial (P^{s}f)}{\partial \zeta } \label{eth}\\
\overline{\eth }f&=&P^{1+s}\frac{\partial (P^{-s}f)}{\partial \overline{\zeta}} \label{ethb}
\end{eqnarray}
where $f$ has a spin weight $s$ and $P$ is the conformal factor defining the sphere metric
\begin{equation}
ds^2=\frac{4d\zeta d\overline \zeta}{P^2}
\end{equation}
note that for axial symmetry, the operators $\eth$ and $\overline{\eth }$ act as derived on $\theta$. In addition, all functions will be functions that do not depend of $\phi$.
In Eqs. (\ref{eth}) and (\ref{ethb}), the conformal factor $P$ is arbitrary. However, in Bondi coordinates, the conformal factor is restricted to
\begin{equation}
P=P_0=1+\zeta \overline{\zeta}
\end{equation}

\subsection{The Newman-Penrose formalism}\label{npnu}
\quad Although the Newman-Penrose (NP) formalism is the basic working tool for our analysis, we will simply give an outline of the formulation and leave the reference \cite{ntod} for details.
We focus in the general form of the asymptotically flat solutions of Einstein-Maxwell equations in Bondi coordinates.

The NP version \cite{ntod},\cite{np} of the vacuum Einstein (or the Einstein-Maxwell) equations uses
the tetrad components
\begin{equation}
{\lambda ^{a}}_{c}=(l^{a}, n^{a}, m^{a}, \overline{m}^{a});\qquad c=1,2,3,4
\end{equation}
rather than the metric, as the basic variable. The metric, Eq. (\ref{metrica}) can be written as
\begin{equation}
g^{ab}=\eta^{cd}{\lambda^{a}}_{c}{\lambda^{b}}_{d}
\end{equation}
with
\begin{equation}
\eta ^{cd}=\left(
\begin{array}{cccc}
0 & 1 & 0 & 0 \\
1 & 0 & 0 & 0 \\
0 & 0 & 0 & -1 \\
0 & 0 & -1 & 0%
\end{array}%
\right)
\end{equation}
The Ricci rotation coefficients ${\gamma^{c}}_{df}$ are defined by \cite{ntod},\cite{np}
\begin{equation}
{\gamma^{c}}_{df}={\lambda^{a}}_{d}{\lambda^{b}}_{f}\nabla _{a}{\lambda^{c}}_{b}
\end{equation}
so
\begin{equation}
\gamma _{cdf}=-\gamma _{dcf}
\end{equation}
the 12 spin coefficients are defined as combinations of the ${\gamma^{c}}_{df}$:
\begin{eqnarray} \label{spincoef}
\alpha =\frac{1}{2}(\gamma _{124}-\gamma _{344}); \quad \lambda =-\gamma _{244}; \quad \kappa =\gamma _{131}\nonumber\\
\beta =\frac{1}{2}(\gamma _{123}-\gamma _{343}); \quad \mu =-\gamma _{243}; \quad \rho =\gamma _{134}\\
\gamma =\frac{1}{2}(\gamma _{122}-\gamma _{342}); \quad \nu =-\gamma _{242}; \quad \sigma =\gamma _{133}\nonumber\\
\varepsilon =\frac{1}{2}(\gamma _{121}-\gamma _{341}); \quad \pi =-\gamma _{241}; \quad \tau =\gamma _{132}\nonumber
\end{eqnarray}

The third basic variable in the NP formalism is the Weyl tensor or, equivalently, the following
five complex tetrad components of the Weyl tensor:
\begin{eqnarray}
\psi _{0}=-C_{abcd}l^{a}m^{b}l^{c}m^{d}; \quad \psi _{1}=-C_{abcd}l^{a}n^{b}l^{c}m^{d} \nonumber\\
\psi _{2}=-\frac{1}{2}(C_{abcd}l^{a}n^{b}l^{c}n^{d}-C_{abcd}l^{a}n^{b}m^{c}{\overline{m}}^{d}) \\
\psi _{3}=C_{abcd}l^{a}n^{b}n^{c}{\overline{m}}^{d}; \quad \psi _{4}=-C_{abcd}n^{a}{\overline{m}}^{b}n^{c}{\overline{m}}^{d} \nonumber
\end{eqnarray}

When an electromagnetic field is present, we include the complex tetrad components of
the Maxwell field
\begin{equation}
\phi _{0}=F_{ab}l^{a}m^{b}; \quad \phi _{1}=\frac{1}{2}F_{ab}(l^{a}n^{b}+m^{a}{\overline{m}}^{b}); \quad \phi _{2}=F_{ab}n^{a}{\overline{m}}^{b},
\end{equation}
into the equations \cite{akn}, \cite{ntod}.

The Peeling theorem of Sachs \cite{sachs} tell us that the behavior of the Weyl scalar and the maxwell tensor is given by:
\begin{eqnarray}
\psi _{0}=\psi _{0}^{0}r^{-5}+O(r^{-6}) \nonumber\\
\psi _{1}=\psi _{1}^{0}r^{-4}+O(r^{-5}) \nonumber\\
\psi _{2}=\psi _{2}^{0}r^{-3}+O(r^{-4}) \nonumber\\
\psi _{3}=\psi _{3}^{0}r^{-2}+O(r^{-3}) \\
\psi _{4}=\psi _{4}^{0}r^{-1}+O(r^{-2}) \nonumber\\
\phi _{0}=\phi _{0}^{0}r^{-3}+O(r^{-4}) \nonumber\\
\phi _{1}=\phi _{1}^{0}r^{-2}+O(r^{-3})\nonumber \\
\phi _{2}=\phi _{2}^{0}r^{-1}+O(r^{-2}) \nonumber
\end{eqnarray}
where the quantities with a zero superscript are function only of ($u_{B}$, $\zeta$, $\overline{\zeta }$). The spin coefficients and metric variables are given as \cite{akn}, \cite{ntod}.
\begin{eqnarray} \label{spincoef-des-r}
\kappa &=&\pi =\varepsilon =0; \qquad \rho =\overline{\rho }; \qquad \tau =\overline{\alpha }+\beta \nonumber\\
\rho  &=&-r^{-1}-\sigma ^{0}\overline{\sigma }^{0}r^{-3}+O(r^{-5}) \nonumber\\
\sigma  &=&\sigma ^{0}r^{-2}+[(\sigma ^{0})^{2}\overline{\sigma }^{0}-\psi
_{0}^{0}/2]r^{-4}+O(r^{-5}) \nonumber\\
\alpha  &=&\alpha ^{0}r^{-1}+O(r^{-2}) \nonumber\\
\beta  &=&\beta ^{0}r^{-1}+O(r^{-2}) \\
\gamma  &=&\gamma ^{0}-\psi _{2}^{0}(2r^{2})^{-1}+O(r^{-3}) \nonumber\\
\mu  &=&\mu ^{0}r^{-1}+O(r^{-2}) \nonumber\\
\lambda  &=&\lambda ^{0}r^{-1}+O(r^{-2}) \nonumber\\
\nu  &=&\nu ^{0}+O(r^{-1}) \nonumber
\end{eqnarray}
where the relationships among the r-independent functions
\begin{eqnarray}
\xi ^{0\zeta }&=&-P_{0}; \qquad \overline{\xi }^{0\zeta }=0; \qquad \xi ^{0\overline{\zeta }}=0; \qquad \overline{\xi }^{0\overline{\zeta }}=-P_{0} \nonumber\\
\alpha ^{0}&=&-\overline{\beta }^{0}=-\frac{\zeta }{2}; \qquad \gamma ^{0}=\nu ^{0}=0; \qquad \omega ^{0}=-\overline{\eth }\sigma ^{0} \nonumber\\
\lambda ^{0}&=&\dot{\overline{\sigma }}^{0}; \qquad \mu ^{0}=U^{0}=-1; \qquad \psi _{4}^{0}=-\ddot{\overline{\sigma }}^{0}; \qquad \psi _{3}^{0}=\eth \dot{\overline{\sigma }}^{0} \nonumber\\
&&\psi _{2}^{0}-\overline{\psi }_{2}^{0}=\overline{\eth }^{2}\sigma ^{0}-\eth^{2}\overline{\sigma }^{0}+\overline{\sigma }^{0}\lambda ^{0}-\sigma ^{0}\overline{\lambda }^{0}
\end{eqnarray}

Finally we return to the choice of the null tetrad. If we start from the rescaled metric $g^{ab\ast }=Z^{\prime2}g^{ab}$ the tangent vector to the generators of $\scri$ is rescaled as $n^{a\ast }=Z^{\prime}n^{a}$. Using this null vector we can define all the other vectors of the new tetrad as
\begin{eqnarray}
{l}_{a}^{\ast} &=&\frac{1}{Z^{\prime }}(l_{a}-\frac{L}{r}\overline{m}_{a}-\frac{\overline{L}}{r}m_{a}+\frac {L\overline{L}}{r^{2}}n_{a}) \\
{n}_{a}^{\ast} &=&\frac{1}{Z^{\prime }}n_{a} \\
{m}_{a}^{\ast} &=&\frac{1}{Z^{\prime }}(m_{a}-\frac{L}{r}n_{a}) \\
{\overline{m}}_{a}^{\ast} &=&\frac{1}{Z^{\prime }}(\overline{m}_{a}-\frac{\overline{L}}{r}n_{a})
\end{eqnarray}
where
$$L(u_{B},\zeta ,\overline{\zeta })=-\frac{\eth_{(u_{B})}T}{\dot{T}}=\eth _{(u)}Z(u,\zeta ,\overline{\zeta })|_{u=T(u_B,\zeta ,\bar{\zeta})} .$$
In the above $\eth_{(u_{B})}$ or $\eth_{(u)}$ means to apply the eth operator keeping $u_B$ or $u$ constant respectively.

From this tetrad we can define the new Weyl scalar \cite {akn}, \cite{arnew}
\begin{equation}
\frac{{\psi }_{1}^{0\ast}(u,\zeta ,\overline{\zeta })}{Z^{\prime 2}}=[\psi _{1}^{0}-3L\psi _{2}^{0}+3L^{2}\psi_{3}^{0}-L^{3}\psi _{2}^{0}](u_B,\zeta ,\overline{\zeta })
\end{equation}
from which we can write the following approximation
\begin{equation}
\psi _{1}^{0\ast}=\psi _{1}^{0}-3L\psi _{2}^{0} \label{psi*}
\end{equation}
if we keep up to linear terms in $Z^{\prime}$ and/or $L$.
\subsection{Evolution equations and physical definitions}
\quad Using the peeling theorem all the radial part of Einstein equations can be integrated leaving only the Bianchi identities at $\scri$ as the unsolved equations. Some of those equations are used to relate Weyl scalars with the free Bondi data, i.e., \cite{ntod}
\begin{eqnarray}
\psi_{2}^{0}-\overline{\psi }_{2}^{0} &=&\overline{\eth }^{2}\sigma ^{0}-\eth ^{2}\overline{\sigma }^{0}+\overline{\sigma }^{0}\dot{\sigma }^{0}-\sigma ^{0}\dot{\overline{\sigma }}^{0}\label {psi2}\\
\psi_{3}^{0}&=&\eth \dot{\overline{\sigma }}^{0}\\
\psi_{4}^{0}&=&-\ddot{\overline{\sigma }}^{0}
\end{eqnarray}
In the above $\sigma^{0}$ is the free data. From Eq. (\ref{psi2}) we can define the so called mass aspect \cite{akn}
\begin{equation} \label{asp.masa}
\Psi =\psi _{2}^{0}+\eth ^{2}\overline{\sigma }^{0}+\sigma ^{0}\dot{\overline{\sigma }^{0}}
\end{equation}
which satisfies the following condition
\begin{equation}
\Psi =\overline{\Psi }
\end{equation}
and finally the evolution equations (Bianchi identities) \cite{ntod}
\begin{eqnarray}
\dot{\psi_{1}^{0}}&=&-\eth \Psi +\eth ^{3}\overline{\sigma }^{0}+\eth \sigma ^{0}\dot{\overline{\sigma }}^{0}+3\sigma ^{0}\eth \dot{\overline{\sigma} }^{0}+\frac{4G}{c^4}\phi _{1}^{0}\overline{\phi }_{2}^{0} \label{psi1prima}\\
\dot{\psi_{2}^{0}}&=&-\eth ^{2}\dot{\overline{\sigma }}^{0}-\sigma ^{0}\ddot{\overline{\sigma }}^{0}+\frac{2G}{c^4}\phi _{2}^{0}\overline{\phi }_{2}^{0} \label{bianchi2}\\
\dot{\phi_{1}^{0}} &=&-\eth \phi _{2}^{0} \\
\dot{\phi_{0}^{0}} &=&-\eth \phi _{1}^{0}+\sigma ^{0}\phi _{2}^{0}
\end{eqnarray}
Using the mass aspect $\Psi$ with $\dot{\psi_{2}^{0}}$, the second of the asymptotic Bianchi identities can be rewritten in the concise form
\begin{eqnarray} \label{psiprima}
\dot{\Psi}=\dot{\sigma} ^{0}\dot{\overline{\sigma }}^{0}+\frac{2G}{c^4}\phi _{2}^{0}\overline{\phi }_{2}^{0}.
\end{eqnarray}
Note that in the above equation the gravitational radiation $\sigma^0$ and the electromagnetic radiation $\phi _{2}^{0}$ determine the mass aspect $\Psi$.
In addition, we can define the Bondi mass and linear momentum as
\begin{eqnarray}
M &=&-\frac{c^{2}}{2\sqrt{2}G}\int \Psi d\Omega  \\
P^{i}&=&-\frac{c^{3}}{6G}\int {\Psi }l^{i}d\Omega,
\end{eqnarray}
and one can easily see that the Bondi mass decreases as a result of the emitted radiation.
\section{Angular Momentum}
The definition of angular momentum in general relativity has proven to be a major task which so far does not have a satisfactory solution. Basically the problem lies at identifying a canonical origin at null infinity. However, for vacuum axially symmetric spacetimes one can use the Komar integral associated with the rotation Killing field $\xi _{(\varphi )}^{a}$ and write a conserved quantity
\begin{equation}\label{komar}
J^{z}=\frac{1}{16\pi }\underset{S_{t}\rightarrow \infty }{\lim}\oint\limits_{S_{t}}\nabla ^{a}\xi _{(\varphi )}^{b}dS_{ab}= const.
\end{equation}
We now want to extend this definition to include the contribution of the electromagnetic radiation.

Using Stokes theorem and the fact that $\xi _{(\varphi )}^{a}$ is a Killing field we have
\begin{equation}
\underset{\partial \Sigma }\oint \nabla ^{a}\xi _{(\varphi )}^{b}dS_{ab}=2\int_{\Sigma }R_{ab}\xi_{(\varphi )}^{b}d\Sigma ^{a}
\end{equation}
where $\partial \Sigma$ is the boundary of the hypersurface $\Sigma $. Since the Killing vector is tangent to $\scri$, ie $\xi _{(\varphi )}^{b}n_{b}=0$, we can replace the Ricci tensor by the stress energy tensor $T_{ab}$, in the above, i.e.,
\begin{equation}
\underset{\partial \Sigma }\oint \nabla ^{a}\xi _{(\varphi )}^{b}dS_{ab}=16\pi \int_{\Sigma }T_{ab}\xi _{(\varphi )}^{b}d\Sigma ^{a}
\end{equation}
Inserting the stress-energy tensor of electromagnetic field  $T_{ab}=\frac{1}{4\pi }(F_{ac}F_{b}^{c}-\frac{1}{4}g_{ab}F^{cd}F_{cd})$ in the r.h.s. of the above equation yields
\begin{eqnarray}
\underset{\partial \Sigma }\oint \nabla ^{a}\xi _{(\varphi )}^{b}dS_{ab}
&=&4\int_{\Sigma }{F_{a}}^{c}F_{bc}\xi _{(\varphi )}^{b}d\Sigma ^{a} \\
&=&4\int_{\Sigma }{F_{a}}^{c}(\nabla _{b}A_{c}-\nabla _{c}A_{b})\xi_{(\varphi )}^{b}d\Sigma ^{a}
\end{eqnarray}
Since we can choose the Maxwell potential to have axial symmetry and the Maxwell field is pure radiation we have in addition
$$\xi _{(\varphi )}^{b}\nabla _{b}A_{c}+A_{b}\nabla_{c}\xi _{(\varphi )}^{b}=0$$
and  $$\nabla ^{c}F_{ac}=0.$$ Thus,
\begin{equation}
\underset{\partial \Sigma }\oint \nabla ^{a}\xi _{(\varphi )}^{b}dS_{ab}=-4\int_{\Sigma }\nabla _{c}(A_{b}\xi _{(\varphi )}^{b}{F_{a}}^{c})d\Sigma
^{a}.
\end{equation}
Using Stokes theorem once again, we finally obtain
\begin{equation}
-4\int_{\Sigma }\nabla _{c}(A_{b}\xi _{(\varphi )}^{b}F^{ac})d\Sigma_{a}=-2\underset{\partial \Sigma }\oint A_{b}\xi _{(\varphi )}^{b}F^{ac}dS_{ac}.
\end{equation}

We thus redefine the angular momentum Eq. (\ref{komar}), to include electromagnetic field, as
\begin{equation} \label{komarfinal}
J^{z}_T=\frac{1}{16\pi }\underset{S_{t}\rightarrow \infty }{\lim}\oint\limits_{S_{t}}[\nabla ^{a}\xi _{(\varphi )}^{b}+2 A_{c}\xi _{(\varphi )}^{c}F^{ab}]dS_{ab}
\end{equation}
The first integrand is the original gravitational term whereas the second one is the electromagnetic contribution. Using the N-P formalism\cite{ntod} one can write Eq. (\ref{komarfinal}) as (see Appendix A),

\begin{equation}\label{J-total}
J_{T}^{z}=\frac{\sqrt{2}c^{3}}{12G}[Im(\psi _{1}^{0}+\sigma ^{0}\eth %
\sigma ^{0})|_{l=1}-4Im(A^{0}\phi _{1}^{0})|_{l=1}]
\end{equation}
where $A^{0}(u_B,\zeta,\bar{\zeta})$ is the Maxwell potential free data related to the electromagnetic radiation via
$$
\overline{\phi _{2}^{0}}= \dot{A}^{0}.
$$

The new conserved quantity $J^{z}_T$ will be called total angular momentum for any axially symmetric Einstein-Maxwell spacetime ie
\begin{equation}
\dot{J}^{z}_T=0
\end{equation}

\section{Center of mass}
\quad Since by assumption the spacetime is axially symmetric we will also assume the center of mass is given by a worldline $R^a(u)$ along the axis of symmetry, i.e., along the \textit{z}-axis. We recall that in section \ref{npnu}, we introduced a null tetrad based on outgoing null hypersurfaces $u=const.$ We will then assume that this family of hypersurfaces has been generated by the future light cones of $R^a(u)$. The intersection of these light cones with $\scri ^+$ yield Newman-Unti coordinates $(u,\zeta,\bar{\zeta})$. The basic idea is to start with the mass dipole term at $\scri$ in a Bondi frame, use Eq. (\ref{psi*}) to write down the transformation equation to a Newman-Unti frame and demand that the mass dipole term vanishes on the $u=const.$ slices. (A similar idea has been used before in the Kozameh-Newman approach for axially symmetric spacetimes\cite{KOR} but as we will see later following both approaches yield different results.)

\subsection{Analysis and definition}
\quad In a Bondi frame, the mass dipole momentum for asymptotically flat spacetime is defined to be the real part of the $l=1$ component of $Re[\psi _{1}^{0}]$. We extend this definition to a Newman-Unti frame and define the mass dipole momentum as the $l=1$ component of  $Re[\psi _{1}^{0\ast}]$.

The basic idea to obtain the center of mass is to start by imposing the condition that on the $u=const$ cuts generated by the worldline $R^a(u)$, the mass dipole momentum vanishes. Then, using the relation (\ref{psi*}) and expanding $\psi _{1}^{0}$ in a tensorial spherical harmonic basis as

$$\psi _{1}^{0}= \psi _{1}^{0i}(u_B) Y^1_{1i}+ \psi _{1}^{0ij}(u_B) Y^1_{2ij}+...$$ one should obtain a relationship between
$\Re[psi _{1}^{0i}(u_B)]$ and the center of mass worldline $\dot{R}^{i}$.

Following this prescription, and using Eq.(\ref{psi*}), on a $u=const.$ slice we impose
\begin{equation}
Re[\psi _{1}^{0}(Z,\zeta,\bar{\zeta})-3\eth(Z) \psi _{2}^{0}(Z,\zeta,\bar{\zeta})]^{i}_{u=const.}=0, \label{psi1uB1}
\end{equation}
 where we have replaced $u_B$ by the function $u_B=Z(R^a(u),\zeta,\bar{\zeta})$. Furthermore, using a slow motion approximation and keeping up to first order terms in the velocity of the center of mass we write
 $$Z(R^a(u),\zeta,\bar{\zeta})=u+\delta u,$$
where we assume $\delta u$ is small. Thus, we make a Taylor expansion of
$$Re[\psi _{1}^{0\ast}(u,\zeta,\bar{\zeta})]=Re[\psi _{1}^{0}(u+\delta u,\zeta,\bar{\zeta})-3\eth\delta u\psi _{2}^{0}(u+\delta u,\zeta,\bar{\zeta})],$$
decompose each term in spherical harmonics and demand that on the $u=const.$ cut the $l=1$ part of this series vanishes. The Taylor expansion yields

$$Re[\psi _{1}^{0\ast}(u,\zeta,\bar{\zeta})]=Re[\psi _{1}^{0}(u,\zeta,\bar{\zeta})+\psi^{0\prime}_{1}(u,\zeta,\bar{\zeta})\delta u-3\eth\delta u \psi _{2}^{0}(u,\zeta,\bar{\zeta})]$$
where we have omitted second order terms in $\delta u$. Taking the $l=1$ part of the above expression and putting it equal to zero yields the following expression,
\begin{eqnarray} \label{RePsi_1}
Re[\psi _{1}^{0}(u)]^{i}=Re[(\eth \Psi-\eth ^{3}\bar{\sigma }^{0})\delta u ]^{i}+3Re[\eth \delta u(\Psi -\eth ^{2}\bar{\sigma}^{0})]^{i},
\end{eqnarray}
i.e., the real, $l=1$ part of $\psi _{1}^{0}$ can be written in terms of the center of mass and other Weyl scalars at null infinity.

Inserting the following tensorial spin-s harmonics expansion \cite{ngilb}
\begin{eqnarray*}
Z &=&u+\delta u \\
\delta u &=&-\frac{1}{2}R^{i}(u)Y_{1i}^{0}(\zeta
)+x^{ij}(u)Y_{2ij}^{0}(\zeta )+x^{ijk}(u)Y_{3ijk}^{0}(\zeta ) \\
\eth \delta u &=&R^{i}(u)Y_{1i}^{1}(\zeta )-6x^{ij}(u)Y_{2ij}^{1}(\zeta
)-12x^{ijk}(u)Y_{3ijk}^{1}(\zeta ) \\
\sigma _{B} &=&\sigma ^{ij}(u_{B})Y_{2ij}^{2}(\zeta )+\sigma
^{ijk}(u_{B})Y_{3ijk}^{2}(\zeta ) \\
\psi _{1}^{0} &=&\psi _{1}^{0i}(u_{B})Y_{1i}^{1}(\zeta )+\psi
_{1}^{0ij}(u_{B})Y_{2ij}^{1}(\zeta )+\psi
_{1}^{0ijk}(u_{B})Y_{3ijk}^{0}(\zeta ) \\
\Psi  &=&-\frac{2\sqrt{2}G}{c^{2}}M(u_{B})-\frac{6G}{c^{3}}%
P^{i}(u_{B})Y_{1i}^{0}(\zeta )+\Psi ^{ij}(u_{B})Y_{2ij}^{0}(\zeta )+\Psi
^{ijk}(u_{B})Y_{3ijk}^{0}(\zeta ) \\
\phi _{0}^{0} &=&\phi _{0}^{0i}(u_{B})Y_{1i}^{1}(\zeta )+\phi
_{0}^{0ij}(u_{B})Y_{2ij}^{1}(\zeta )\\
\phi _{1}^{0} &=&Q(u_{B})+\phi _{1}^{0i}(u_{B})Y_{1i}^{0}(\zeta )+\phi
_{1ij}^{0}(u_{B})Y_{2ij}^{0}(\zeta )\\
\phi _{2}^{0} &=&\phi _{2}^{0i}(u_{B})Y_{1i}^{-1}(\zeta )+\phi
_{2}^{0ij}(u_{B})Y_{2ij}^{-1}(\zeta )
\end{eqnarray*}
in Eq.(\ref{RePsi_1}) yields
\begin{eqnarray}\label{Di}
D^{i} &=&MR^{i}-\frac{96}{5\sqrt{2}c}x^{ij}P^{j}+%
\frac{3c^{2}}{5\sqrt{2}G}R^{j}\Psi ^{ij} \nonumber\\
&&+\frac{c^{2}}{7\sqrt{2}G}[x^{jk}(72\sigma _{R}^{ijk}-216\Psi ^{ijk})+x^{ijk}(432\Psi ^{jk} -360\sigma
_{R}^{jk})],
\end{eqnarray}
where we have defined
$$
D^{i}(u)\equiv -\frac{c^{2}}{6\sqrt{2}G}Re[\psi _{1}^{0}(u)]^i
$$
(Since we assume axial symmetry, $R^{i}(u)$ only has a $\textit{z}$ component. Likewise, all the high order tensors are symmetric, diagonal and trace-free.)

We now take a small digression to concentrate on the light cone cut function $Z=u+\delta u$ defined as the intersection of the future lightcone from a worldline $x^a(u)$ with $\scri ^+$. The function $Z$ dynamically depends on the matter and radiation content of the spacetime via the solution of the Einstein equations, i.e., the light cone cut function is dynamical variable and we do not make any a priori assumption about its behaviour. $Z$ satisfies the equation

$$\eth^2 Z= \Lambda(Z,\eth Z, \bar{\eth Z}, \eth \bar{\eth Z}, \zeta, \bar{\zeta}),$$
and $\Lambda$ satisfies the Einstein's equations (it vanishes for a flat spacetime). The freedom in the solution is given by a combination of $l=0,1$ spherical harmonics since they are annihilated by the $\eth^2$ operator.

One can write this freedom as
$$Z_0= t(u) +x^{i}(u)Y_{1i}^{0}(\zeta ,\bar{\zeta})= u+x^{i}(u)Y_{1i}^{0},$$
where int he last equality we have thrown away quadratic and higher order terms in $v^{i}(u)$. The function $\Lambda$ only contains $l=2$ and higher spherical harmonics decomposition which are completely determined from the Einstein's equations. For example, if we assume a vacuum spacetime in the neighborhood of null infinity, the linearized equation for $\Lambda$ is given by
\begin{equation} \label{ethethbZ}
\bar{\eth }^{2}\Lambda=\eth ^{2}\bar{\sigma }_B(Z_0, \zeta ,\bar{\zeta})+\bar{\eth }^{2}\sigma_B(Z_0, \zeta,\bar{\zeta}).
\end{equation}
It follows from the above equation that given any point $x^a$ of the spacetime, the $l=2$ and higher terms of $Z$ are completely determined from $\sigma_B(u, \zeta,\bar{\zeta})$. For example, up to linear order terms in $\sigma_B$ and/or $x^i$ we have
\begin{eqnarray}\label{Xij}
x^{ij} &=&\frac{1}{12}\sigma _{R}^{ij} \\
x^{ijk} &=&\frac{1}{60}\sigma _{R}^{ijk},\label{Xijk}
\end{eqnarray}
where the subscript $R$ means the real part of the complex quantities. (If we keep bilinear terms of $\sigma_B$ and $x^i$ in $\ref{ethethbZ}$, then the above terms also depend on $x^i(u)$.) Inserting $x^{ij}$ and $x^{ij}$ in Eq.(\ref{RePsi_1}) gives an explicit relationship between $x^i(u)$ and $Re[\psi _{1}^{0}(u)]^{i}$. For the particular assumption given in (\ref{Xij}), (\ref{Xijk}) we get
\begin{equation}
D^{i}=MR^{i}-\frac{8}{5\sqrt{2}c}\sigma _{R}^{ij}P^{j}+\frac{c^{2}}{\sqrt{2}G}[\frac{3}{5}R^{j}\psi ^{ij}-\frac{18}{7}\sigma _{R}^{jk}\psi ^{ijk}+\frac{36}{35}\sigma _{R}^{ijk}\psi ^{jk}]
\end{equation}
It follows from the Bianchi identities that the time evolution of $\Psi$ is quadratic in $\dot{\sigma} _B$. Assuming $\sigma _B$, $\Psi ^{ij}$, and $\Psi ^{ijk}$ vanish at $u_B= - \infty$ and keeping up to second order terms in $\sigma _B$ or $R^{i}$ we get the following expression
\begin{eqnarray} \label{psi1real}
D^{i} + \frac{8}{5\sqrt{2}c}\sigma _{R}^{ij}P^{j} =M R^{i}.
\end{eqnarray}
 We have thus obtained an explicit relationship between the coordinate of the center of mass $R^{i}$ and the Weyl scalars defined at Null Infinity. Even if we assume a more involved field equation for $Z$ giving a functional dependence of $x^{ij}$ and $x^{ijk}$ on $R^{i}$ as well as on $\sigma_B$, Eq.(\ref{RePsi_1}) will give an algebraic expression relating $R^{i}$ with the Weyl scalars at null infinity from which one can solve for the center of mass worldline.

 To obtain a relation between the velocity of the center of mass and the Bondi linear momentum we take a time derivative of (\ref{psi1real}). Using again the Bianchi identities  yields the following expression
\begin{eqnarray} \label{Pz}
P^{i}+\frac{4}{5}\dot{\sigma}_{R}^{ij}P^{j}-\frac{3c^{3}}{14G}(\sigma ^{ijk}%
\dot{{\overline{\sigma }}}^{jk}-\sigma ^{jk}\dot{{\overline{\sigma }}}%
^{ijk})_{R}-\frac{1}{3c}(\phi _{1}^{0}\overline{\phi }_{2}^{0})_{R}^{i}=%
\frac{c\sqrt{2}}{2}MV^{i}
\end{eqnarray}
(In the above expression we have omitted the terms $\dot{M}$, $\dot{P^i}$, $\dot{\Psi}^{ij}$, $\dot{\Psi}^{ijk}$ since they are quadratic in $\sigma_B$ and $\phi_{2}^{0}$.)
Since all the vector quantities are aligned with the symmetry axis and the tensor variables are symmetric and trace free, we can get a simplified form of the equations as

\begin{eqnarray}
D^{z}+\frac{8}{5\sqrt{2}c}\sigma _{R}^{zz}P^{z}&=&MR^{z} \\
P^{z}+\frac{4}{5}\dot{\sigma}_{R}^{zz}P^{z}-\frac{3c^{3}}{14G}(\sigma ^{zjk}%
\dot{{\overline{\sigma }}}^{jk}-\sigma ^{jk}\dot{{\overline{\sigma }}}%
^{zjk})_{R}-\frac{1}{3c}(\phi _{1}^{0}\overline{\phi }_{2}^{0})_{R}^{z}&=&%
\frac{c\sqrt{2}}{2}MV^{z}\label{Pz}
\end{eqnarray}
These equations provide explicit relations between $D^{z}$, $P^{z}$ and $R^{z}$, $V^{z}$.

The equation for the linear momentum can also be written as a sum of different parts as

\begin{equation}
P^{z}=P_M^{z}+P_G^{z}+P_{EM}^{z},
\end{equation}

with
\begin{eqnarray}
P_M^{z}&=&(1-\frac{4}{5}\dot{\sigma}_{R}^{zz})\frac{c\sqrt{2}}{2}MV^{z}, \\
P_G^{z}&=&\frac{3c^{3}}{14G}(\sigma ^{zjk}\dot{{\bar{\sigma }}}^{jk}-\sigma ^{jk}\dot{{\bar{\sigma }}}^{zjk})_{R}, \\
P_{EM}^{z}&=&\frac{1}{3c}(\phi _{1}^{0}\bar{\phi }_{2}^{0})_{R}^{z},\label{dipole}
\end{eqnarray}
emphasizing the role of each contribution to the total linear momentum. The leading term in (\ref{dipole}) is proportional to the charge times the time derivative of the electric dipole contribution. If one further assumes that the dipole contribution is due to a charged particle with worldline $R^i$ one recovers a known result, the Abraham-Lorentz momentum\cite{lorentz-abrahams}. However we are here concerned with astrophysical compact objects and this term usually vanishes.
\subsection{Equation of motion}

Taking a time derivative of (\ref{Pz}) and inserting the Biachi identity
\begin{eqnarray}
\dot{P}^{z}&=&-\frac{c^{3}}{6G}[\frac{3}{7}(\dot{\sigma }^{zjk}\dot{\bar{\sigma }}^{jk}+\dot{\sigma }^{jk}\dot{\bar{\sigma }}^{zjk})+\frac{2G}{c^{4}}(\phi_{2}^{0}\bar{\phi }_{2}^{0})^{z}]\label{ppuntom}\\
\end{eqnarray}
gives the equation of motion for the center of mass,
\begin{eqnarray}\label{Az}
M (\dot{V}^{z}- \frac{4}{5}\ddot{\sigma}_{R}^{zz} V^{z}) &=& -\frac{\sqrt{2}c^2}{7G}[\frac{3}{2}(\sigma
^{zjk}\dot{{\bar{\sigma }}}^{jk}-\sigma ^{jk}\dot{{\overline{\sigma }}}%
^{zjk}\dot{)}_{R}+(\dot{\sigma }^{zjk}\dot{\bar{\sigma }}^{jk})_R] \\
&&- \frac{\sqrt{2}}{3c^2}[(\phi_{2}^{0}\bar{\phi }_{2}^{0})^{z}+(\phi _{1}^{0}\overline{\phi }_{2}^{0}\dot{)}_{R}^{z}]. \nonumber
\end{eqnarray}
For completeness we also give the mass loss equation
\begin{equation}\label{massloss}
\dot{M}=-\frac{c^{2}}{2\sqrt{2}G}[\frac{1}{5}\dot{\sigma}^{ij}\dot{\overline{%
\sigma }}^{ij}+\frac{6}{7}\dot{\sigma}^{ijk}\dot{\overline{\sigma }}^{ijk}]-%
\frac{\sqrt{2}}{6c^{2}}\overline{\phi }_{2}^{0i}\phi _{2}^{0i}
\end{equation}
From the r.h.s. of (\ref{Az}) we define the notion of gravitational and electromagnetic forces, i.e.,
\begin{eqnarray}\label{Forces}
F_G &\equiv& -\frac{\sqrt{2}c^2}{7G}[\frac{3}{2}(\sigma
^{zjk}\dot{{\bar{\sigma }}}^{jk}-\sigma ^{jk}\dot{{\bar{\sigma }}}%
^{zjk}\dot{)}_{R}+(\dot{\sigma }^{zjk}\dot{\bar{\sigma }}^{jk})_R]\\
F_{EM} &\equiv& - \frac{\sqrt{2}}{3c^2}[(\phi_{2}^{0}\bar{\phi }_{2}^{0})^{z}+(\phi _{1}^{0}\overline{\phi }_{2}^{0}\dot{)}_{R}^{z}].
\end{eqnarray}

Note that the gravitational force vanishes when the quadrupole or octupole moment vanish. In this case the electromagnetic radiation will produce the acceleration of the center of mass. This acceleration, however, will be negligible for most situations. Since the mass loss equation has a separate contribution from the quadrupole and octupole moment, the net effect in this situation will be a reduction of the gravitational mass of the system while the center of mass remains at rest. On the other hand, most head-on collisions between compact objects will produce quadrupole and octupole radiation terms and there will be a net acceleration of the center of mass.

It also follows from Eq. (\ref{Az}) that there are no runaway solutions. The functions $\ddot{\sigma}_{R}^{zz}$, $F_G$ and $F_{EM}$ decrease to zero as $u \to \infty$. Asymptotically this equation gives a constant velocity if the total radiation is finite. Thus, the motion of the center of mass does not have runaway behavior.

\section{Applications}
\quad In this section we will first check that the formalism developed gives the correct answer for the cases where definitions have already been given. We will also compare our work with others to analyze similarities and differences in the definitions of linear and angular momentum.  We begin with applying this formalism to the case of a stationary and axially symmetric spacetime.
\subsection{Stationary and axially symmetric spacetime}
\quad First consider the Kerr metric, in this case we have $\psi _{4}^{0}=\psi_{3}^{0}=0$. Moreover as the spacetime is stationary all derivatives respect $u_B$ vanish. In this way we would have to $\dot{\sigma}=0$, so $\sigma =\sigma(\zeta, \bar{\zeta})$. Without loss of generality, we will work in a referential where $\sigma=0$. From Eq. (\ref{psiprima}) we get that $P^i=\Psi^{ij}=\Psi^{ijk}=0$ for all $i, j, k$ and from Eq. (\ref{asp.masa}) we get that $\psi_{2}^{0} \propto M$. Furthermore form Eq. (\ref{ethethbZ}) is easy to show that $x ^{ij}=x ^{ijk}=0$.
\begin{equation} \label{psi1kerr}
\psi _{1}^{0z}=\frac{6G}{c^{2}}M\left( -{\sqrt{2}R}^{z}{+i\frac{2}{\sqrt{2}c}a}\right)
\end{equation}
where $a$ is the angular parameter. The real part of $\psi _{1}^{0z}$ is given from the Eq. (\ref{psi1real}), so from Eq. (\ref{J-total}) we have
\begin{equation}
J^{z}_{T}=aM
\end{equation}
which corresponds to the angular momentum of the Kerr spacetime. If now we consider the Kerr-Newman case, the development is very similar to the Kerr case. Note that in  Eq. (\ref{J-total}) $A_{c}m^{c}=0$, so we get the same equations
\begin{eqnarray}
\psi _{1}^{0z}&=&\frac{6G}{c^{2}}M\left( -{\sqrt{2}R}^{z}{+i\frac{2}{\sqrt{2}c}a}\right) \\
P^x&=&P^y=P^z=0 \\
J^{z}_{T}&=&aM
\end{eqnarray}
note that although these equations are identical, the evolution of the mass center is different for Kerr-Newman spacetime due to the presence of electromagnetic fields.

\subsection{Massive explosions or head on collisions.}
\quad We consider here either a massive explosion of an isolated system, like type I supernova, or a massive head-on collision. We assume that initially the center of mass is at rest. Immediately after the explosion or collision the acceleration of the center of mass will be given by
\begin{eqnarray*}
M \dot{V}^{z} &=& -\frac{\sqrt{2}c^2}{7G}[\frac{3}{2}(\sigma
^{zjk}\dot{{\bar{\sigma }}}^{jk}-\sigma ^{jk}\dot{{\overline{\sigma }}}%
^{zjk}\dot{)}_{R}+(\dot{\sigma }^{zjk}\dot{\bar{\sigma }}^{jk})_R] \\
&&- \frac{\sqrt{2}}{3c^2}[(\phi_{2}^{0}\bar{\phi }_{2}^{0})^{z}+(\phi _{1}^{0}\overline{\phi_{2}^{0}}\dot{)}_{R}^{z}]. \nonumber
\end{eqnarray*}

Note that if either the quadrupole or octupole term vanishes there is no gravitational contribution to the acceleration. Any collision will have a quadrupole term but only collisions between uneven masses will also have an octupole contribution.  Likewise, the electromagnetic force will be dominated by the radiation term since for most astrophysical objects $\phi _{1}^{0}=0$.

Although total angular momentum is conserved, the coupling  between gravitational and electromagnetic angular momentum gives a transfer mechanism by which the system can gain or loose intrinsic angular momentum. Consider for simplicity that initially the system does not have angular momentum. After the explosion or collision the system will acquire an intrinsic gravitational angular momentum if electromagnetic radiation is emitted, i.e., from
\begin{equation*}
J_{T}^{z}=J_{G}^{z}-\frac{\sqrt{2}c^{3}}{12G}[4Im(A^{0}\phi _{1}^{0})|_{l=1}]=0.
\end{equation*}
the electromagnetic angular momentum creates an intrinsic gravitational angular momentum in the opposite direction of the electromagnetic one. This effect could be important in charged isolated systems like the positron cloud discovered by the COMPTON detector in GRO, but will be negligible for most cases.

\subsection{Comparison with AKN equations}
\quad In this subsection we will compare our equations with the Adamo-Kozameh-Newman (AKN) equations for linear and angular momentum. For simplicity we will consider a vacuum spacetime in the neighborhood of null infinity and assume the gravitational radiation only has quadrupole terms. Directly from \cite{akn} we write
\begin{equation}
P^{z}=\frac{\sqrt{2}}{2}McV^{z}-\frac{9}{10}\frac{M}{c}(\dot{V}^{z} \sigma_{R}^{zz}+V^{z} \dot{\sigma}_{R}^{zz})-\frac{1}{10}\frac{M}{c}(\ddot{\xi}_{I}^{z}\sigma _{I}^{zz}+\dot{\xi}_{I}^{z}\dot{\sigma} _{I}^{zz})-\frac{6}{10}\frac{c^{2}}{G}(2R^{z}\sigma _{R}^{zz}+\xi _{I}^{z}\sigma _{I}^{zz}\dot{)}
\end{equation}
where the scalar $\xi _{I}^{z}$ is related to the intrinsic angular momentum via
$$
S^{z} \equiv \frac{\sqrt{2}}{2}Mc\xi _{I}^{z}
$$

Likewise, The total angular momentum in the AKN formalism is given by
\begin{equation}\label{J-AKN}
J_T^{z}=\frac{\sqrt{2}}{2}Mc\xi _{I}^{z}-\frac{3}{10}\frac{c^{3}}{G}(\xi _{I}^{z}\sigma_{R}^{zz}-2R^{z}\sigma _{I}^{zz})-\frac{9}{10}MV^{z}\sigma _{I}^{zz}+\frac{1}{20}M\dot{\xi}_{I}^{z}\sigma _{R}^{zz}.
\end{equation}

In this work the equivalent equations are given by
\begin{equation}
P^{z}=\frac{c\sqrt{2}}{2}MV^{z}(1 - \frac{4}{5}\dot{\sigma}_{R}^{zz})
\end{equation}
and
\begin{equation}
J_T^{z}=S^{z}= const.,
\end{equation}
as one can see from applying the Komar approach and using a Bondi and a Newman-Unti cut as the boundary $\partial \Sigma$.

Although both formulations agree for stationary spacetimes, they differ when gravitational radiation is present. It follows from the Bianchi identities and the Komar integral for the axially symmetric Killing field that either in this or the AKN formulation we have

$$P^{z}=const.$$
$$J_T^{z}= const.$$

Thus, in this formulation $S^{z}= const.$ and $V^{z}$ decreases to an asymptotic value after the gravitational radiation is emitted. In the AKN formulation both $V^{z}$ and $S^{z}$ are functions of time and obey a coupled system of ODEs. They also decay to an asymptotic value when the gravitational radiation is emitted.

Also in this formulation the relationship between total and intrinsic angular momentum appears to be natural. Since the orbital part of the angular momentum vanishes when the center of mass vector $R^i$ and the velocity $V^i$ are aligned along the $z$-axis, one expects that the intrinsic and total angular momentum should be equal. In the AKN formulation, the intrinsic angular momentum depends on the position and velocity of the center of mass and one does not expect this kind of relation. However, naturalness is not easy to define when gravitational radiation is present and it may well be that a more involved relationship of the form given in (\ref{J-AKN}) is correct. A nice test for the two formulations will be available when gravitational wave astronomy is finally developed.

\section{Conclusions}

Using the available geometric structure of asymptotically flat spacetimes together with conservation laws that arise when those spacetimes are axially symmetric, we have defined the notion of linear and angular momentum for Einstein Maxwell spaces.

Furthermore, using the light cone equation we have been able to identify worldlines inside the spacetime with Newman Unti cuts at null infinfity. The center of mass worldline $R^a$ is then selected by imposing the condition that the mass dipole moment at null infinity vanishes when restricted to the center of mass NU cut. Using the available Bianchi identities at $\scri ^+$ one obtains a relationship between the center of mass velocity and the Bondi momentum as well as the equation of motion of $R^a$.

Several nice highlights of this approach are
\begin{itemize}
\item A definition of angular momentum when electromagnetic fields are present.
\item A definition of center of mass worldline and velocity which are algebraically related to radiation fields at null infinity.
\item Definitions of gravitational and electromagnetic forces in terms of radiation fields.
\item Appropriate behaviour of the equations of motion (no runaway solutions).
\item A natural relationship between intrinsic and total angular momentum (they are the same in this case).
\end{itemize}

The equations of motion could be used in astrophysical situations when the system has axial symmetry to predict the motion of the center of mass if the radiation is detected or to predict the amount of radiation if the velocity and acceleration of the center of mass is available.

The formalism is ready to be generalized for spaces without symmetries and it will be considered in future work. In the generalization we expect some new features that are absent in axially symmetric spaces. Since at the moment there is no definition of angular momentum that has been universally accepted one can either work with a parameter dependent definition or use a suitable radiation data where all the definitions agree. It is left for the future to find a new definition of angular momentum for any kind of gravitational radiation.
Acknowledgements: this research has been supported by grants from CONICET and the Agencia Nacional de Ciencia y Tecnología.

\appendix
\section{Komar integral and angular momentum}
In vacuum spaces the Komar integral of the Killing field $\xi _{(\varphi )}^{b}$ yields a definition of the \textit{z}-component of the angular momentum,
\begin{equation}\label{intkomar}
J^{z}=\frac{1}{16\pi }\underset{S_{t}\rightarrow \infty }{\lim}\oint\limits_{S_{t}}\nabla ^{a}\xi _{(\varphi )}^{b}dS_{ab}
\end{equation}
One can explicitly integrate this equation in the N-P formalism to obtain a formula at $\scri$ in terms of the spin coefficients.  We first write the Killing vector field $\xi _{(\varphi )}^{b}$ as a combination of the null tetrad vectors as
\begin{equation}\label{killing}
\xi _{(\varphi )}^{b}=\xi_{l}l^{b}+\bar{\xi_m} m^{b}+\xi_m \bar{m} ^{b}+\xi_{n} n^{b}
\end{equation}
where
\begin{eqnarray}
\xi_{l} &=& \frac{Im(\sigma^{0}\bar{\omega}^{0})}{r}\cos \theta\\
\xi_m&=& ir\cos \theta\\
\xi_{n}&=&0
\end{eqnarray}
and the two-dimensional surface area can also be expressed as
\begin{equation}
dS_{ab}=-2n_{[a}l_{b]}r^{2}\sin \theta  d\theta  d\varphi,
\end{equation}
thus the Komar integral can be written as
\begin{equation}
J^{z}=-\frac{1}{16\pi }\underset{r\rightarrow \infty }{\lim }%
\int\nolimits_{0}^{\pi }\int\nolimits_{0}^{2\pi }\nabla ^{a}\xi _{(\varphi )}^{b}(n_{a}l_{b}-l_{a}n_{b})r^{2}\sin \theta d\varphi d\theta.
\end{equation}

Using Eqs. (\ref{spincoef}), (\ref{spincoef-des-r}) and Eqs. (\ref{ln}),(\ref{mmb}) and writing this equation up to order $O(r^{-2})$ we get
\begin{equation}
\nabla ^{a}\xi _{(\varphi )}^{b}(n_{a}l_{b}-l_{a}n_{b})=\frac{\cos \theta }{r^{2}}Im(\psi _{1}^{0}-\sigma ^{0}\bar{\omega }^{0})
\end{equation}
where $\bar{\omega }^{0}=-\eth \bar{\sigma }^{0}$ \cite{ntod},\cite{nu}. Thus, the Komar integral can be written as
\begin{equation}
J^{z}=\frac{1}{8}\int\nolimits_{0}^{\pi }Im(\psi _{1}^{0}+\sigma ^{0}\eth \bar{\sigma }^{0})\cos \theta d(\cos \theta ).
\end{equation}
where we have used the axial symmetry to integrate in the azimuth direction. Finally, this integral gives the following definition of angular momentum \cite{szab}
\begin{equation} \label{momang}
J^{z} \propto Im(\psi _{1}^{0}+\sigma ^{0}\eth \bar{\sigma }^{0})|_{l=1}
\end{equation}

We can follow a similar calculation with Eq. (\ref{komarfinal})
\begin{equation}
J^{z}_T=\frac{1}{16\pi }\underset{S_{t}\rightarrow \infty }{\lim}\oint\limits_{S_{t}}[\nabla ^{a}\xi _{(\varphi )}^{b}+2 A_{c}\xi _{(\varphi )}^{c}F^{ab}]dS_{ab}
\end{equation}
Using the fact that \cite{ntod}
\begin{equation}
F^{ab}=2\phi _{0}\bar{m}^{[a}n^{b]}+\phi _{1}(n^{[a}l^{b]}+m^{[a}\bar{m}^{b]})+2\phi _{2}l^{[a}m^{b]}
\end{equation}
the second integral can be put in the form
\begin{equation}
\frac{1}{16\pi }\underset{S_{t}\rightarrow \infty }{\lim}\oint\limits_{S_{t}} 2 A_{c}\xi _{(\varphi )}^{c}F^{ab}dS_{ab}=\frac{1}{8}\underset{r\rightarrow \infty }{\lim} \int\limits_{0}^{\pi}2A_{c}\xi _{(\varphi )}^{c}\phi _{1}^{0}\sin \theta d\theta
\end{equation}
so, we can define the total angular momentum as
\begin{equation} \label{J-total}
J^{z}_T= \frac{\sqrt{2}c^3}{12G} [Im(\psi _{1}^{0}+\sigma ^{0}\eth \bar{\sigma }^{0})|_{l=1}+\underset{r\rightarrow \infty }{\lim}\int\limits_{0}^{\pi}2A_{c}\xi _{(\varphi )}^{c}\phi _{1}^{0}\sin \theta d\theta]
\end{equation}
Using the tetrad decomposition of the vector killing field $\xi _{(\varphi )}^{b}$ at $\scri +$, and the fact that $A_c$ and $\phi_1^0$ are real, one can rewrite $J_{EM}$ as
\begin{eqnarray*}
J_{EM} &=&\underset{r\rightarrow \infty }{\lim }\int_{0}^{\pi }2A_{c}\xi
_{(\varphi )}^{c}\phi _{1}^{0}\sin \theta d\theta  \\
&=&-\underset{r\rightarrow \infty }{\lim }\int_{0}^{\pi }2A_{c}\xi
_{(\varphi )}^{c}\phi _{1}^{0}d(\cos \theta ) \\
&=&-\underset{r\rightarrow \infty }{\lim }\int_{0}^{\pi }2irA_{c}(\overline{m%
}^{c}-m^{c})\phi _{1}^{0}\cos \theta d(\cos \theta ) \\
&=&-4\underset{r\rightarrow \infty }{\lim }r\int_{0}^{\pi }Im%
(A_{c}m^{c})\phi _{1}^{0}\cos \theta d(\cos \theta ) \\
&=&-4\underset{r\rightarrow \infty }{\lim }rIm(A_{c}m^{c}\phi
_{1}^{0})|_{l=1}
\end{eqnarray*}

Furthermore, it can be shown that,
$$
\underset{r\rightarrow \infty }{\lim }rA_{c}m^{c}=A^{0},
$$
where $A^{0}(u_B,\zeta,\bar{\zeta})$ is the free Maxwell potential data related to the electromagnetic radiation via
$$
\overline{\phi _{2}^{0}}= \dot{A}^{0}.
$$
Thus, the total angular momentum is finally expressed as
\begin{equation*}
J_{T}^{z}=\frac{\sqrt{2}c^{3}}{12G}[Im(\psi _{1}^{0}+\sigma ^{0}\eth %
\sigma ^{0})|_{l=1}-4Im(A_0 \phi _{1}^{0})|_{l=1}]
\end{equation*}
\section{Tensorial spin-s harmonics products}
We present a table of tensorial harmonics products which complete the list of product \cite{ngilb}.\\

{\bf Products of the form $Y_{1i}^{s}Y_{3jkl}^{s}$}
\begin{eqnarray}
Y_{1i}^{-1}Y_{3jkl}^{2}&=&\frac{5}{21}F_{ijkl}^{2(1)}-\frac{2}{21}G_{ijkl}^{2(1)}-\frac{1}{24}i\sqrt{2}F_{ijkl}^{3(1)}-\frac{1}{56}F_{ijkl}^{4(1)}\\
Y_{1i}^{0}Y_{3jkl}^{1}&=&\frac{20}{21}F_{ijkl}^{2(1)}-\frac{8}{21}G_{ijkl}^{2(1)}-\frac{1}{12}i\sqrt{2}F_{ijkl}^{3(1)}+\frac{5}{28}F_{ijkl}^{4(1)}\\
Y_{1i}^{1}Y_{3jkl}^{0}&=&-\frac{10}{7}F_{ijkl}^{2(1)}+\frac{4}{7}G_{ijkl}^{2(1)}+\frac{1}{2}i\sqrt{2}F_{ijkl}^{3(1)}+\frac{5}{14}F_{ijkl}^{4(1)}\\
Y_{1i}^{0}Y_{3jkl}^{0}&=&\frac{10}{7}F_{ijkl}^{2(0)}-\frac{4}{7}G_{ijkl}^{2(0)}+\frac{1}{7}F_{ijkl}^{4(0)}\\
Y_{1i}^{-1}Y_{3jkl}^{1}&=&\frac{5}{21}F_{ijkl}^{2(0)}-\frac{2}{21}G_{ijkl}^{2(0)}+\frac{1}{24}i\sqrt{2}F_{ijkl}^{3(0)}-\frac{1}{56}F_{ijkl}^{4(0)}\\
Y_{1i}^{1}Y_{3jkl}^{-1}&=&\frac{5}{21}F_{ijkl}^{2(0)}-\frac{2}{21}G_{ijkl}^{2(0)}-\frac{1}{24}i\sqrt{2}F_{ijkl}^{3(0)}-\frac{1}{56}F_{ijkl}^{4(0)}
\end{eqnarray}
where
\begin{eqnarray*}
F_{ijkl}^{2(s)}&=&\delta _{ij}Y_{2kl}^{s}+\delta _{ik}Y_{2jl}^{s}+\delta
_{il}Y_{2jk}^{s}\\
G_{ijkl}^{2(s)}&=&\delta _{jk}Y_{2il}^{s}+\delta _{kl}Y_{2ij}^{s}+\delta
_{jl}Y_{2ik}^{s}\\
F_{ijkl}^{3(s)}&=&\epsilon _{ijm}Y_{3klm}^{s}+\epsilon
_{ikm}Y_{3jlm}^{s}+\epsilon _{ilm}Y_{3jkm}^{s}\\
F_{ijkl}^{4(s)}&=&Y_{4ijkl}^{s}
\end{eqnarray*}
with the superscript $s=0, 1$.\\

{\bf Products of the form $Y_{2ij}^{s}Y_{3klm}^{s}$}
\begin{eqnarray}
Y_{2ij}^{1}Y_{3klm}^{0}&=&-\frac{1}{12}F_{ijklm}^{1(1)}+\frac{24}{35}G_{ijklm}^{1(1)}+\frac{24}{35}H_{ijklm}^{1(1)}-\frac{2}{7}i\sqrt{2}F_{ijklm}^{2(1)}+\frac{5}{7}i\sqrt{2}G_{ijklm}^{2(1)}+\nonumber\\&& +\frac{2}{15}F_{ijklm}^{3(1)}+\frac{1}{15}G_{ijklm}^{3(1)}+\frac{1}{15}H_{ijklm}^{3(1)}+\frac{1}{14}i\sqrt{2}F_{ijklm}^{4(1)}+\frac{2}{21}F_{ijklm}^{5(1)}\\
Y_{2ij}^{0}Y_{3klm}^{1}&=&\frac{1}{12}F_{ijklm}^{1(1)}-\frac{24}{35}G_{ijklm}^{1(1)}-\frac{24}{35}H_{ijklm}^{1(1)}+\frac{1}{7}i\sqrt{2}F_{ijklm}^{2(1)}-\frac{5}{14}i\sqrt{2}G_{ijklm}^{2(1)}+\nonumber\\&& +\frac{1}{5}F_{ijklm}^{3(1)}+\frac{1}{10}G_{ijklm}^{3(1)}+\frac{1}{10}H_{ijklm}^{3(1)}-\frac{1}{28}i\sqrt{2}F_{ijklm}^{4(1)}+\frac{1}{14}F_{ijklm}^{5(1)}\\
Y_{2ij}^{-2}Y_{3klm}^{3}&=&\frac{1}{14}F_{ijklm}^{1(1)}-\frac{1}{35}G_{ijklm}^{1(1)}-\frac{1}{35}H_{ijklm}^{1(1)}-\frac{1}{168}i\sqrt{2}F_{ijklm}^{2(1)}+\nonumber\\&& +\frac{5}{336}i\sqrt{2}G_{ijklm}^{2(1)}-\frac{1}{180}F_{ijklm}^{3(1)}-\frac{1}{360}G_{ijklm}^{3(1)}-\frac{1}{360}H_{ijklm}^{3(1)}+\\&&
-\frac{1}{1680}i\sqrt{2}F_{ijklm}^{4(1)}+\frac{1}{5040}F_{ijklm}^{5(1)}\nonumber\\
Y_{2ij}^{2}Y_{3klm}^{-1}&=&-\frac{1}{7}F_{ijklm}^{1(1)}+\frac{2}{35}G_{ijklm}^{1(1)}+\frac{2}{35}H_{ijklm}^{1(1)}-\frac{1}{28}i\sqrt{2}F_{ijklm}^{2(1)}+\nonumber\\&& +\frac{5}{56}i\sqrt{2}G_{ijklm}^{2(1)}+\frac{1}{15}F_{ijklm}^{3(1)}+\frac{1}{30}G_{ijklm}^{3(1)}+\frac{1}{30}H_{ijklm}^{3(1)}-\\&&
-\frac{1}{84}i\sqrt{2}F_{ijklm}^{4(1)}-\frac{1}{168}F_{ijklm}^{5(1)}\nonumber\\
Y_{2ij}^{-1}Y_{3klm}^{2}&=&\frac{2}{7}F_{ijklm}^{1(1)}-\frac{4}{35}G_{ijklm}^{1(1)}-\frac{4}{35}H_{ijklm}^{1(1)}+\frac{1}{30}F_{ijklm}^{3(1)}+\frac{1}{60}G_{ijklm}^{3(1)}+\nonumber\\&& +\frac{1}{60}H_{ijklm}^{3(1)}+\frac{1}{120}i\sqrt{2}F_{ijklm}^{4(1)}-\frac{1}{210}F_{ijklm}^{5(1)}\\
Y_{2ij}^{0}Y_{3klm}^{0}&=&\frac{36}{7}F_{ijklm}^{1(0)}-\frac{72}{35}G_{ijklm}^{1(0)}-\frac{72}{35}H_{ijklm}^{1(0)}+\frac{4}{15}F_{ijklm}^{3(0)}+\frac{2}{15}G_{ijklm}^{3(0)}+\nonumber\\&& +\frac{2}{15}H_{ijklm}^{3(0)}+\frac{1}{21}F_{ijklm}^{5(0)}\\
Y_{2ij}^{1}Y_{3klm}^{-1}&=&\frac{4}{7}F_{ijklm}^{1(0)}-\frac{8}{35}G_{ijklm}^{1(0)}-\frac{8}{35}H_{ijklm}^{1(0)}+\frac{1}{42}i\sqrt{2}F_{ijklm}^{2(0)}+\nonumber\\&&
-\frac{5}{84}i\sqrt{2}G_{ijklm}^{2(0)}+\frac{1}{90}F_{ijklm}^{3(0)}+\frac{1}{180}G_{ijklm}^{3(0)}+\frac{1}{180}H_{ijklm}^{3(0)}-\\&&
-\frac{1}{168}i\sqrt{2}F_{ijklm}^{4(0)}-\frac{1}{252}F_{ijklm}^{5(0)}\nonumber\\
Y_{2ij}^{-1}Y_{3klm}^{1}&=&\frac{4}{7}F_{ijklm}^{1(0)}-\frac{8}{35}G_{ijklm}^{1(0)}-\frac{8}{35}H_{ijklm}^{1(0)}-\frac{1}{42}i\sqrt{2}F_{ijklm}^{2(0)}+\nonumber\\&&
+\frac{5}{84}i\sqrt{2}G_{ijklm}^{2(0)}+\frac{1}{90}F_{ijklm}^{3(0)}+\frac{1}{180}G_{ijklm}^{3(0)}+\frac{1}{180}H_{ijklm}^{3(0)}-\\&&
+\frac{1}{168}i\sqrt{2}F_{ijklm}^{4(0)}-\frac{1}{252}F_{ijklm}^{5(0)}\nonumber
\end{eqnarray}
\begin{eqnarray}
Y_{2ij}^{-2}Y_{3klm}^{2}&=&\frac{1}{14}F_{ijklm}^{1(0)}-\frac{1}{35}G_{ijklm}^{1(0)}-\frac{1}{35}H_{ijklm}^{1(0)}-\frac{1}{168}i\sqrt{2}F_{ijklm}^{2(0)}+\nonumber\\&&
+\frac{5}{336}i\sqrt{2}G_{ijklm}^{2(0)}-\frac{1}{180}F_{ijklm}^{3(0)}-\frac{1}{360}G_{ijklm}^{3(0)}-\frac{1}{360}H_{ijklm}^{3(0)}-\\&&
-\frac{1}{1680}i\sqrt{2}F_{ijklm}^{4(0)}+\frac{1}{5040}F_{ijklm}^{5(0)}\nonumber\\
Y_{2ij}^{2}Y_{3klm}^{-2}&=&\frac{1}{14}F_{ijklm}^{1(0)}-\frac{1}{35}G_{ijklm}^{1(0)}-\frac{1}{35}H_{ijklm}^{1(0)}+\frac{1}{168}i\sqrt{2}F_{ijklm}^{2(0)}-\nonumber\\&&
-\frac{5}{336}i\sqrt{2}G_{ijklm}^{2(0)}-\frac{1}{180}F_{ijklm}^{3(0)}-\frac{1}{360}G_{ijklm}^{3(0)}-\frac{1}{360}H_{ijklm}^{3(0)}+\\&&
+\frac{1}{1680}i\sqrt{2}F_{ijklm}^{4(0)}+\frac{1}{5040}F_{ijklm}^{5(0)}\nonumber
\end{eqnarray}
where (with $s=0, 1$)
\begin{eqnarray*}
F_{ijklm}^{1(s)}&=&(\delta _{ik}\delta _{jl}+\delta _{il}\delta_{jk})Y_{1m}^{s}+(\delta _{il}\delta _{jm}+\delta _{im}\delta_{jl})Y_{1k}^{s}+(\delta _{im}\delta _{jk}+\delta _{ik}\delta_{jm})Y_{1l}^{s}\\
G_{ijklm}^{1(s)}&=&(\delta _{jk}\delta _{lm}+\delta _{jl}\delta _{mk}+\delta_{jm}\delta _{kl})Y_{1i}^{s}+(\delta _{ik}\delta _{lm}+\delta _{il}\delta_{mk}+\delta _{im}\delta _{kl})Y_{1j}^{s}\\
H_{ijklm}^{1(s)}&=&\delta _{ij}(\delta _{kl}Y_{1m}^{s}+\delta_{lm}Y_{1k}^{s}+\delta _{mk}Y_{1l}^{s})\\
F_{ijklm}^{2(s)}&=&(\delta _{lm}\epsilon _{ikf}+\delta _{km}\epsilon_{ilf}+\delta _{lk}\epsilon _{imf})Y_{2jf}^{s}+(\delta _{lm}\epsilon_{jkf}+\delta _{km}\epsilon _{jlf}+\delta _{lk}\epsilon _{imf})Y_{2if}^{s}\\
G_{ijklm}^{2(s)}&=&\delta _{il}(\epsilon _{jmf}Y_{2kf}^{s}+\epsilon_{jkf}Y_{2mf}^{s})+\delta _{im}(\epsilon _{jkf}Y_{2lf}^{s}+\epsilon_{jlf}Y_{2kf}^{s})+\\&&
+\delta _{ik}(\epsilon _{jlf}Y_{2mf}^{s}+\epsilon_{jmf}Y_{2lf}^{s})+\delta _{jl}(\epsilon _{imf}Y_{2kf}^{s}+\epsilon_{ikf}Y_{2mf}^{s})+\\&&
+\delta _{jm}(\epsilon _{ikf}Y_{2lf}^{s}+\epsilon_{ilf}Y_{2kf}^{s})+\delta _{jk}(\epsilon _{ilf}Y_{2mf}^{s}+\epsilon_{imf}Y_{2lf}^{s})\\
F_{ijklm}^{3(s)}&=&\delta _{ij}Y_{3klm}^{s}\\
G_{ijklm}^{3(s)}&=&\delta _{jm}Y_{3ikl}^{s}+\delta _{jk}Y_{3ilm}^{s}+\delta_{jl}Y_{3ikm}^{s}+\delta _{im}Y_{3jkl}^{s}+\delta _{ik}Y_{3jlm}^{s}+\delta_{il}Y_{3jkm}^{s}\\
H_{ijklm}^{3(s)}&=&(\epsilon _{ikf}\epsilon _{jln}+\epsilon _{jkf}\epsilon_{iln}+\epsilon _{ilf}\epsilon _{jkn}+\epsilon _{jlf}\epsilon_{ikn})Y_{3fnm}^{s}+\\&&
+(\epsilon _{ilf}\epsilon _{jmn}+\epsilon _{jlf}\epsilon_{imn}+\epsilon _{imf}\epsilon _{jln}+\epsilon _{jmf}\epsilon_{iln})Y_{3fnk}^{s}+\\&&
+(\epsilon _{imf}\epsilon _{jkn}+\epsilon _{jmf}\epsilon _{ikn}+\epsilon_{ikf}\epsilon _{jmn}+\epsilon _{kmf}\epsilon _{imn})Y_{3fnl}^{s}\\
F_{ijklm}^{4(s)}&=&\epsilon _{ikf}Y_{4fjlm}^{s}+\epsilon_{ilf}Y_{4fjkm}^{s}+\epsilon _{imf}Y_{4fjkl}^{s}+\epsilon_{jkf}Y_{4film}^{s}+\epsilon _{jlf}Y_{4fikm}^{s}+\epsilon_{jmf}Y_{4fikl}^{s} \\
F_{ijklm}^{5(s)}&=&Y_{5ijklm}^{s}
\end{eqnarray*}
{\bf Products of the form $Y_{3ijk}^{s}Y_{3lmn}^{s}$}
\begin{eqnarray}
Y_{3ijk}^{0}Y_{3lmn}^{1}&=&\frac{6}{7}i\sqrt{2}F_{ijklmn}^{1(1)}-\frac{30}{7}i\sqrt{2}G_{ijklmn}^{1(1)}+\frac{1}{3}F_{ijklmn}^{2(1)}+\frac{1}{3}G_{ijklmn}^{2(1)}-\\&&
-\frac{2}{21}H_{ijklmn}^{2(1)}+\frac{9}{7}K_{ijklmn}^{2(1)}-\frac{5}{18}i\sqrt{2}F_{ijklmn}^{3(1)}+\frac{2}{9}i\sqrt{2}G_{ijklmn}^{3(1)}+\nonumber\\&&
+\frac{2}{9}i\sqrt{2}H_{ijklmn}^{3(1)}+\frac{15}{154}F_{ijklmn}^{4(1)}+\frac{15}{154}G_{ijklmn}^{4(1)}+\frac{25}{154}H_{ijklmn}^{4(1)}-\nonumber\\&&
-\frac{5}{252}i\sqrt{2}F_{ijklmn}^{5(1)}+\frac{5}{132}F_{ijklmn}^{6(1)}\nonumber\\
Y_{3ijk}^{-2}Y_{3lmn}^{3}&=&\frac{1}{140}i\sqrt{2}F_{ijklmn}^{1(1)}-\frac{1}{28}i\sqrt{2}G_{ijklmn}^{1(1)}+\frac{1}{72}F_{ijklmn}^{2(1)}+\frac{1}{72}G_{ijklmn}^{2(1)}-\\&&
-\frac{1}{252}H_{ijklmn}^{2(1)}+\frac{3}{56}K_{ijklmn}^{2(1)}+\frac{1}{216}i\sqrt{2}F_{ijklmn}^{3(1)}-\frac{1}{270}i\sqrt{2}G_{ijklmn}^{3(1)}+\nonumber\\&&
-\frac{1}{270}i\sqrt{2}H_{ijklmn}^{3(1)}-\frac{1}{616}F_{ijklmn}^{4(1)}-\frac{1}{616}G_{ijklmn}^{4(1)}-\frac{5}{1848}H_{ijklmn}^{4(1)}-\nonumber\\&&
-\frac{1}{6048}i\sqrt{2}F_{ijklmn}^{5(1)}+\frac{1}{15840}F_{ijklmn}^{6(1)}\nonumber
\end{eqnarray}
\begin{eqnarray}
Y_{3ijk}^{2}Y_{3lmn}^{-1}&=&-\frac{1}{14}i\sqrt{2}F_{ijklmn}^{1(1)}+\frac{5}{14}i\sqrt{2}G_{ijklmn}^{1(1)}+\frac{1}{12}F_{ijklmn}^{2(1)}+\frac{1}{12}G_{ijklmn}^{2(1)}-\\&&
-\frac{1}{42}H_{ijklmn}^{2(1)}+\frac{9}{28}K_{ijklmn}^{2(1)}+\frac{1}{77}F_{ijklmn}^{4(1)}+\frac{1}{77}G_{ijklmn}^{4(1)}+\nonumber\\&&
+\frac{5}{231}H_{ijklmn}^{4(1)}-\frac{1}{336}i\sqrt{2}F_{ijklmn}^{5(1)}-\frac{1}{528}F_{ijklmn}^{6(1)}\nonumber\\
Y_{3ijk}^{-1}Y_{3lmn}^{2}&=&\frac{1}{14}i\sqrt{2}F_{ijklmn}^{1(1)}-\frac{5}{14}i\sqrt{2}G_{ijklmn}^{1(1)}+\frac{1}{12}F_{ijklmn}^{2(1)}+\frac{1}{12}G_{ijklmn}^{2(1)}-\nonumber\\&&
-\frac{1}{42}H_{ijklmn}^{2(1)}+\frac{9}{28}K_{ijklmn}^{2(1)}+\frac{1}{77}F_{ijklmn}^{4(1)}+\frac{1}{77}G_{ijklmn}^{4(1)}+\\&&
+\frac{5}{231}H_{ijklmn}^{4(1)}+\frac{1}{336}i\sqrt{2}F_{ijklmn}^{5(1)}-\frac{1}{528}F_{ijklmn}^{6(1)}\nonumber\\
Y_{3ijk}^{0}Y_{3lmn}^{0}&=&-\frac{48}{7}F_{ijklmn}^{0}+\frac{120}{7}G_{ijklmn}^{0}+\frac{4}{3}F_{ijklmn}^{2(0)}+\frac{4}{3}G_{ijklmn}^{2(0)}-\nonumber\\&&
-\frac{8}{21}H_{ijklmn}^{2(0)}+\frac{36}{7}K_{ijklmn}^{2(0)}+\frac{9}{77}F_{ijklmn}^{4(0)}+\\&&
+\frac{9}{77}G_{ijklmn}^{4(0)}+\frac{15}{77}H_{ijklmn}^{4(0)}+\frac{5}{231}F_{ijklmn}^{6(0)}\nonumber\\
Y_{3ijk}^{-1}Y_{3lmn}^{1}&=&-\frac{4}{7}F_{ijklmn}^{0}+\frac{10}{7}G_{ijklmn}^{0}-\frac{1}{14}i\sqrt{2}F_{ijklmn}^{1(0)}+\frac{5}{14}i\sqrt{2}G_{ijklmn}^{1(0)}+\nonumber\\&&
+\frac{1}{12}F_{ijklmn}^{2(0)}+\frac{1}{12}G_{ijklmn}^{2(0)}-\frac{1}{42}H_{ijklmn}^{2(0)}+\frac{9}{28}K_{ijklmn}^{2(0)}+\\&&
+\frac{5}{216}i\sqrt{2}F_{ijklmn}^{3(0)}-\frac{1}{54}i\sqrt{2}G_{ijklmn}^{3(0)}-\frac{1}{54}i\sqrt{2}H_{ijklmn}^{3(0)}+\nonumber\\&&
+\frac{1}{616}F_{ijklmn}^{4(0)}+\frac{1}{616}G_{ijklmn}^{4(0)}+\frac{5}{1848}H_{ijklmn}^{4(0)}-\nonumber\\&&
-\frac{5}{3024}i\sqrt{2}F_{ijklmn}^{5(0)}-\frac{5}{3696}F_{ijklmn}^{6(0)}\nonumber\\
Y_{3ijk}^{-2}Y_{3lmn}^{2}&=&-\frac{2}{35}F_{ijklmn}^{0}+\frac{1}{7}G_{ijklmn}^{0}-\frac{1}{70}i\sqrt{2}F_{ijklmn}^{1(0)}+\frac{1}{14}i\sqrt{2}G_{ijklmn}^{1(0)}+\nonumber\\&&
+\frac{1}{432}i\sqrt{2}F_{ijklmn}^{3(0)}-\frac{1}{540}i\sqrt{2}G_{ijklmn}^{3(0)}-\frac{1}{540}i\sqrt{2}H_{ijklmn}^{3(0)}-\\&&
-\frac{1}{880}F_{ijklmn}^{4(0)}-\frac{1}{880}G_{ijklmn}^{4(0)}-\frac{1}{528}H_{ijklmn}^{4(0)}-\nonumber\\&&
-\frac{1}{7560}i\sqrt{2}F_{ijklmn}^{5(0)}+\frac{1}{18480}F_{ijklmn}^{6(0)}\nonumber
\end{eqnarray}
where
\begin{eqnarray*}
F_{ijklmn}^{0} &=&\delta _{ij}(\delta _{kl}\delta _{mn}+\delta _{km}\delta_{nl}+\delta _{kn}\delta _{lm})+\delta _{jk}(\delta _{il}\delta _{mn}+\delta_{im}\delta _{nl}+\delta _{in}\delta _{lm})+ \\&&
+\delta _{ki}(\delta _{jl}\delta _{mn}+\delta _{jm}\delta _{nl}+\delta_{jn}\delta _{lm})\\
G_{ijklmn}^{0}&=&\delta _{il}(\delta _{jm}\delta _{kn}+\delta _{jn}\delta_{km})+\delta _{jl}(\delta _{km}\delta _{in}+\delta _{kn}\delta_{im})+\delta _{kl}(\delta _{im}\delta _{jn}+\delta_{in}\delta _{jm})\\
F_{ijklmn}^{1(s)}&=&(\delta _{ij}\delta _{lm}\epsilon _{knf}+\delta_{ij}\delta _{ln}\epsilon _{kmf}+\delta _{ij}\delta _{nm}\epsilon_{klf}+\delta _{jk}\delta _{lm}\epsilon _{inf}+\delta_{jk}\delta_{ln}\epsilon_{imf}+\\&&
+\delta _{jk}\delta_{nm}\epsilon_{ilf}+\delta _{ki}\delta _{lm}\epsilon _{jnf}+\delta _{ki}\delta _{ln}\epsilon_{jmf}+\delta _{ki}\delta _{nm}\epsilon _{jlf})Y_{1f}^{s}\\
G_{ijklmn}^{1(s)} &=&(\delta _{il}\delta _{jm}\epsilon _{knf}+\delta_{im}\delta _{jn}\epsilon _{klf}+\delta _{in}\delta _{jl}\epsilon_{kmf}+\delta _{jl}\delta _{km}\epsilon _{inf}+\delta _{jm}\delta_{kn}\epsilon _{ilf}+ \\&&
+\delta _{jn}\delta _{kl}\epsilon _{imf}+\delta _{kl}\delta _{im}\epsilon _{jnf}+\delta _{km}\delta _{in}\epsilon_{jlf}+\delta _{kn}\delta _{il}\epsilon _{jmf})Y_{1f}^{s}\\
F_{ijklmn}^{2(s)} &=&(\delta _{mn}\delta _{kl}+\delta _{ml}\delta_{kn}+\delta _{ln}\delta _{km})Y_{2ij}^{s}+(\delta _{mn}\delta _{il}+\delta_{ml}\delta _{in}+\delta _{ln}\delta _{im})Y_{2jk}^{s}+\\&&
+(\delta _{mn}\delta _{jl}+\delta _{ml}\delta _{jn}+\delta _{ln}\delta_{jm})Y_{2ik}^{s}\\
G_{ijklmn}^{2(s)}&=&(\delta _{ij}\delta _{kl}+\delta _{jk}\delta_{il}+\delta _{ik}\delta _{jl})Y_{2mn}^{s}+(\delta _{ij}\delta _{km}+\delta_{jk}\delta _{im}+\delta _{ik}\delta _{jm})Y_{2ln }^{s}+ \\
&&+(\delta _{ij}\delta _{kn}+\delta _{jk}\delta _{in}+\delta _{ik}\delta_{jn})Y_{2lm}^{s}\\
\end{eqnarray*}
\begin{eqnarray*}
H_{ijklmn}^{2(s)} &=&(\delta _{jk}\delta _{mn}+\delta _{jm}\delta_{kn}+\delta _{jn}\delta _{km})Y_{2il}^{s}+(\delta _{jk}\delta _{ln }+\delta_{jn}\delta _{kl}+\delta _{jl}\delta _{kn})Y_{2im}^{s}+\\&&
+(\delta _{jk}\delta_{lm}+\delta _{jl}\delta _{km}+\delta _{jm}\delta _{kl})Y_{2in}^{s}+(\delta _{ik}\delta _{mn}+\delta _{im}\delta _{kn}+\delta _{in}\delta_{km})Y_{2jl}^{s}+\\&&
+(\delta _{ik}\delta _{ln }+\delta _{in}\delta _{kl}+\delta_{il}\delta _{kn})Y_{2jm}^{s}+(\delta _{ik}\delta _{lm}+\delta _{il}\delta_{km}+\delta _{im}\delta _{kl})Y_{2jn}^{s}+ \\
&&+(\delta _{ji}\delta _{mn}+\delta _{jm}\delta _{in}+\delta _{jn}\delta_{im})Y_{2kl}^{s}+(\delta _{ji}\delta _{ln }+\delta _{jn}\delta _{il}+\delta_{jl}\delta _{in})Y_{2km}^{s}+\\&&
+(\delta _{ji}\delta _{lm}+\delta _{jl}\delta_{im}+\delta _{jm}\delta _{il})Y_{2kn}^{s}\\
K_{ijklmn}^{2(s)} &=&(\epsilon _{ilf}\epsilon _{jmg}\delta _{kn}+\epsilon_{imf}\epsilon _{jng}\delta _{kl}+\epsilon _{inf}\epsilon _{jlg}\delta_{km}+\epsilon _{jlf}\epsilon _{kmg}\delta_{in}+\epsilon _{jmf}\epsilon_{kng}\delta _{il}+\\&&
+\epsilon _{jnf}\epsilon _{klg}\delta _{im}+\epsilon _{klf}\epsilon _{img}\delta _{jn}+\epsilon _{kmf}\epsilon_{ing}\delta _{jl}+\epsilon _{knf}\epsilon _{ilg}\delta _{jm})Y_{2fg}^{s}\\
F_{ijklmn}^{3(s)} &=&\delta _{il}(\epsilon _{jmf}Y_{3fkn}^{s}+\epsilon_{jnf}Y_{3fkm}^{s}+\epsilon _{kmf}Y_{3fjn}^{s}+\epsilon_{knf}Y_{3fjm}^{s})+\delta _{im}(\epsilon_{jnf}Y_{3fkl}^{s}+\\&&
\epsilon_{jlf}Y_{3fkn}^{s}+\epsilon _{knf}Y_{3fjl}^{s}+\epsilon _{klf}Y_{3fjn}^{s})+\delta _{in}(\epsilon _{jlf}Y_{3fkm}^{s}+\epsilon_{jmf}Y_{3fkl}^{s}+\epsilon_{klf}Y_{3fjm}^{s}+\\&&
+\epsilon_{kmf}Y_{3fjl}^{s})+\delta _{jl}(\epsilon _{kmf}Y_{3fin}^{s}+\epsilon_{knf}Y_{3fim}^{s}+\epsilon _{imf}Y_{3fkn}^{s}+\epsilon _{inf }Y_{3fkm}^{s})+\\&&
+\delta _{jm}(\epsilon_{knf}Y_{3fil}^{s}+\epsilon_{klf}Y_{3fin}^{s}+\epsilon _{inf }Y_{3fkl}^{s}+\epsilon_{ilf}Y_{3fkn}^{s})+\delta _{jn}(\epsilon_{klf}Y_{3fim}^{s}+\\&&
+\epsilon_{kmf}Y_{3fil}^{s}+\epsilon _{ilf}Y_{3fkm}^{s}+\epsilon _{imf}Y_{3fkl}^{s})+\delta _{kl}(\epsilon _{imf}Y_{3fjn}^{s}+\epsilon _{inf}Y_{3fjm}^{s}+\\&&
+\epsilon_{jmf}Y_{3fin}^{s}+\epsilon_{jnf}Y_{3fim}^{s})+\delta _{km}(\epsilon _{inf }Y_{3fjl}^{s}+\epsilon_{ilf}Y_{3fjn}^{s}+\epsilon _{jnf}Y_{3fil}^{s})+\\&&
+\epsilon _{jlf}Y_{3fin}^{s}+\delta_{kn}(\epsilon_{ilf}Y_{3fjm}^{s}+\epsilon_{imf}Y_{3fjl}^{s}+\epsilon _{jlf}Y_{3fim}^{s}+\epsilon _{jmf}Y_{3fil}^{s})\\
G_{ijklmn}^{3(s)} &=&\delta _{ij}(\epsilon _{klf}Y_{3fmn}^{s}+\epsilon_{kmf}Y_{3fln}^{s}+\epsilon_{knf}Y_{3flm}^{s})+\delta_{ik}(\epsilon_{jlf}Y_{3fmn}^{s}+\epsilon _{jmf}Y_{3fln}^{s}+\\&&
+\epsilon _{jnf}Y_{3flm}^{s})+\delta _{jk}(\epsilon _{ilf}Y_{3fmn}^{s}+\epsilon_{imf}Y_{3fln}^{s}+\epsilon _{inf}Y_{3flm}^{s})\\
H_{ijklmn}^{3(s)}&=&\delta _{lm}(\epsilon _{inf}Y_{3fjk}^{s}+\epsilon_{jnf}Y_{3fik}^{s}+\epsilon_{knf}Y_{3fij}^{s})+\delta _{ln}(\epsilon_{imf}Y_{3fjk}^{s}+\epsilon _{jmf}Y_{3fik}^{s}+\\&&
+\epsilon _{kmf}Y_{3fij}^{s})+\delta _{nm}(\epsilon _{ilf}Y_{3fjk}^{s}+\epsilon_{jlf}Y_{3fik}^{s}+\epsilon _{klf}Y_{3fij}^{s})\\
F_{ijklmn}^{4(s)}&=&\delta _{ij}Y_{4klmn}^{s}+\delta _{kj}Y_{4ilmn}^{s}+\delta_{ki}Y_{4jlmn}^{s}\\
G_{ijklmn}^{4(s)}&=&\delta _{lm}Y_{4nijk}^{s}+\delta _{mn}Y_{4lijk}^{s}+\delta_{nl}Y_{4mijk}^{s}\\
H_{ijklmn}^{4(s)}&=&\epsilon _{ilf}\epsilon_{jmg}Y_{4fgkn}^{s}+\epsilon_{jlf}\epsilon_{kmg}Y_{4fgin}^{s}+\epsilon_{klf}\epsilon_{img}Y_{4fgjn}^{s}+\\&&
+\epsilon _{imf}\epsilon _{jng}Y_{4fgkl}^{s}+\epsilon_{jmf}\epsilon _{kng}Y_{4fgil}^{s}+\epsilon_{kmf}\epsilon_{ing}Y_{4fgjl}^{s}+\\&&
+\epsilon _{inf}\epsilon _{jlg}Y_{4fgkm}^{s}+\epsilon _{jnf}\epsilon_{klg}Y_{4fgim}^{s}+\epsilon _{knf}\epsilon_{ilg}Y_{4fgjm}^{s}\\
F_{ijklmn}^{5(s)}&=&\epsilon _{ilf}Y_{5fjkmn}^{s}+\epsilon_{jlf}Y_{5fikmn}^{s}+\epsilon _{klf}Y_{5fijmn}^{s}+\epsilon_{imf}Y_{5fjkln}^{s}+\epsilon _{inf }Y_{5fjklm}^{s}+ \\
&&+\epsilon _{jmf}Y_{5fikln }^{s}+\epsilon _{jnf}Y_{5fiklm}^{s}+\epsilon _{kmf}Y_{5fijln}^{s}+\epsilon_{knf}Y_{5fijlm}^{s}\\
F_{ijklmn}^{6(s)}&=&Y_{6ijklmn}^{s}
\end{eqnarray*}


\end{document}